\documentclass[journal]{IEEEtran}

\usepackage{wrapfig}
\usepackage{booktabs}
\usepackage{algorithm}
\usepackage{algorithmic}
\usepackage{amsmath}
\usepackage{xcolor}
\usepackage{caption}
\usepackage{pifont}
\usepackage{multirow}
\usepackage[numbers,sort&compress]{natbib}
\usepackage[colorlinks, citecolor=blue]{hyperref}
\definecolor{deepblue}{rgb}{0, 0, 0.9}
\definecolor{lightgray}{rgb}{0.4, 0.4, 0.4}
\usepackage{arydshln}
\usepackage{graphicx}
\usepackage{wrapfig}
\usepackage{xcolor}         
\usepackage{hyperref}
\usepackage{subcaption}
\usepackage{url}            
\usepackage{amsfonts}       
\usepackage{nicefrac}       
\usepackage{microtype}      
\usepackage{tikz}

\usepackage{textpos}
\usepackage{pifont}
\usepackage{eso-pic}
\usepackage{atbegshi}
\usepackage{picture}

\makeatletter
\def\ps@IEEEtitlepagestyle{%
    \def\@oddfoot{}%
    \def\@evenfoot{}%
    \def\@oddhead{%
        \vbox{%
            \hbox to\textwidth{%
                \includegraphics[height=0.7cm]{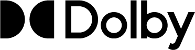}%
                \hfil%
            }%
            \vspace{0.1cm}%
            \hbox to\textwidth{\leaders\hrule height 0.4pt\hfill}
        }%
    }%
    \def\@evenhead{\@oldhead}%
}
\makeatother

\begin{document}
\title{Why Your Tokenizer Fails in Information Fusion: A Timing-Aware Pre-Quantization Fusion for Video-Enhanced Audio Tokenization}



\author{Xiangyu Zhang~\IEEEmembership{Student Member,~IEEE,}, Benjamin John Southwell, Siqi Pan, Xinlei Niu, Beena Ahmed~\IEEEmembership{Member,~IEEE,}, Julien Epps~\IEEEmembership{Senior Member,~IEEE}
\thanks{Xiangyu Zhang, Beena Ahmed, Julien Epps are with the School of Electrical Engineering and Telecommunications, University of New South Wales, Sydney, Australia.}
\thanks{ Xiangyu Zhang, Xinlei Niu, Benjamin John Southwell, Siqi Pan are with Dolby Laboratories, Sydney, Australia}
}


\maketitle

\begin{abstract}
Audio tokenization has emerged as a critical component in end-to-end audio language models, enabling efficient discrete representation learning for both audio understanding and generation tasks. However, existing audio tokenizers face fundamental limitations in understanding tasks due to single-modality constraints, particularly when audio signals contain ambiguous or incomplete information. While incorporating additional modality information can significantly enhance audio understanding, current multimodal fusion approaches invariably degrade reconstruction quality. This degradation is unacceptable for end-to-end audio systems that require high-fidelity audio generation capabilities. In this work, we investigate the root causes of reconstruction quality degradation in video-enhanced audio tokenization and present three key findings. First, the location of fusion within the tokenizer architecture is crucial for preserving reconstruction quality. Second, we show that contrastive learning, though effective in continuous representation fusion, is unsuitable for discrete tokenizers as it fails to enhance downstream task performance. Third, while feature-dimension fusion approaches achieve moderate success, we discover that fusing along the temporal axis—guided by the concept of distinctive features—yields significantly better results. Building on these insights, we introduce the Timing-Aware Pre-Quantization Fusion for Video-Enhanced Audio Tokenization, the first approach to successfully integrate visual information into audio tokenizer architectures while preserving reconstruction fidelity. Our approach not only maintains high-fidelity reconstruction but also achieves superior performance on downstream understanding tasks compared with audio-only tokenizers and established multimodal fusion baselines.
\end{abstract}


\section{Introduction}

The emergence of end-to-end audio language models has revolutionized the landscape of audio processing, enabling systems that can simultaneously understand and generate audio content through unified architectures~\cite{defossez2024moshi,huang2025step,wu2025step,du2024cosyvoice,zhang2024speaking,chen2025selective,tian2025step,wu2025mind}. At the heart of these systems lies audio tokenization—the process of converting continuous audio signals into discrete token sequences that can be efficiently processed by transformer-based architectures~\cite{zeghidour2021soundstream,defossez2022high,kumar2023high,ji2024wavtokenizer}. This discrete representation paradigm offers significant advantages over continuous approaches in computational efficiency, memory usage, and seamless integration with existing language model frameworks~\cite{achiam2023gpt,dubey2024llama}. Most importantly for audio-to-audio applications, discrete representations enable straightforward autoregressive generation through standard language modeling objectives, contrasting with continuous representation systems that require complex generation mechanisms and often struggle with controllable synthesis~\cite{xu2025qwen2,wu2025step}. However, existing audio tokenizers face fundamental limitations that constrain their effectiveness in complex understanding tasks. Audio signals, by their very nature, often contain ambiguous or incomplete information that can lead to multiple valid interpretations~\cite{li2025reinforcement,rouditchenko2025omni}. Consider scenarios where environmental sounds overlap, audio quality is degraded, or contextual information is crucial for accurate interpretation—in such cases, relying solely on audio modality significantly limits the system's understanding capabilities~\cite{10109890,4066991,5332301,10948343,10814093}. This single-modality constraint becomes particularly problematic in real-world applications where audio and visual information naturally co-occur and provide complementary semantic cues.

The integration of visual information presents a compelling solution to these limitations. In natural environments, audio and visual modalities exhibit strong temporal and semantic correspondences that humans exploit effortlessly for enhanced perception~\cite{meredith1986visual,knopfel2019audio}. Visual context can disambiguate ambiguous audio signals, provide semantic enrichment for audio understanding, and offer robust cues when audio information is degraded or incomplete~\cite{knopfel2019audio}. Recent advances in multimodal learning have demonstrated the effectiveness of audio-visual fusion across various continuous representation tasks, validating the potential of cross-modal information integration~\cite{tsiamas2025sequential,wang2024v2a,chen2025selective}.

Despite this promising foundation, extending multimodal fusion to discrete audio tokenization presents unprecedented challenges. A critical requirement for end-to-end audio language models is the ability to generate high-fidelity audio outputs, as reconstruction quality directly impacts the naturalness and intelligibility of synthesized audio~\cite{xu2025qwen2,wu2025step,du2024cosyvoice}. However, existing approaches to incorporating additional information into audio tokenizers consistently exhibit a troubling pattern: while semantic understanding may be enhanced, reconstruction audio quality invariably degrades~\cite{zhang2024speechtokenizer,zhang2025distinctive,gong2025xy,jiang2025unicodec}. This degradation creates a fundamental tension between the dual objectives of multimodal understanding and high-quality audio generation, forcing practitioners to choose between enhanced comprehension and fidelity preservation.

The root causes of this reconstruction quality degradation remain poorly understood. Previous work has primarily focused on adapting fusion strategies from continuous representation learning without adequately considering the unique constraints and objectives of discrete tokenization~\cite{tsiamas2025sequential,wang2024v2a}. The quantization process inherent in tokenization introduces optimization challenges that may not be present in continuous representation spaces, potentially requiring fundamentally different fusion approaches~\cite{huh2023straightening,amjad2019learning}. Furthermore, the temporal nature of both audio and visual modalities suggests that conventional feature-space fusion methods may not optimally exploit the natural correspondences between these modalities~\cite{wang2024tiva}.

To address these fundamental challenges, this work presents a comprehensive investigation into the design principles governing successful video-enhanced audio tokenization. Rather than accepting reconstruction degradation as an inevitable consequence of multimodal integration, we systematically analyze the underlying causes and propose novel solutions. Our investigation reveals three critical insights that challenge conventional approaches to multimodal tokenization. First, the location of fusion within the tokenizer architecture plays a crucial role in determining both reconstruction quality and understanding performance. Through systematic experimentation, we identify optimal fusion points that preserve reconstruction fidelity while enabling effective cross-modal integration. Second, fusion methodologies successful in continuous representation learning do not necessarily translate to discrete tokenizer architectures. Specifically, we find that contrastive learning, despite its effectiveness in continuous multimodal representation learning~\cite{tsiamas2025sequential,li2022blip,li2023blip}, fails to provide meaningful improvements for downstream tasks in discrete tokenization settings while potentially interfering with quantization objectives.Third, temporal-dimension fusion strategies, guided by principles from distinctive feature theory~\cite{lee1988automatic,liu1996landmark,zhang25_interspeech,zhang2025speecht}, significantly outperform conventional feature-dimension fusion approaches. This finding suggests that exploiting the natural temporal correspondences between audio and visual modalities is essential for effective multimodal discrete representation learning.

Building on these insights, we introduce Timing-Aware Pre-Quantization Fusion (TAPF), a novel approach specifically designed for video-enhanced audio tokenization. Our method successfully bridges the gap between multimodal understanding capabilities and reconstruction quality requirements, achieving superior performance on downstream audio understanding tasks while maintaining high-fidelity audio reconstruction. Through extensive experiments on standard benchmarks, we demonstrate that TAPF outperforms both single-modality audio tokenizers and existing multimodal fusion baselines, establishing a new paradigm for discrete multimodal representation learning.


\section{Preliminary}
\label{sec:preliminary}
\subsection{Multimodal Input Representations and Omni-Modal Architectures}

Modern multimodal large language models (MLLMs) process information from multiple sensory modalities through two primary representation paradigms: continuous and discrete inputs~\cite{bai2025qwen2,chu2024qwen2,zhang2023speechgpt}. Continuous representations, such as dense feature vectors extracted from vision encoders~\cite{radford2021learning,liu2021swin} and audio encoders~\cite{radford2023robust}, offer rich semantic information and have proven highly effective for multimodal understanding tasks~\cite{xu2025qwen2,wu2025step}. These representations preserve fine-grained details and enable smooth interpolation between different semantic concepts, making them particularly suitable for tasks requiring nuanced cross-modal reasoning.

In contrast, discrete representations utilize tokenized inputs where each modality is converted into sequences of discrete tokens, similar to text processing in traditional language models~\cite{sennrich2016neural,ahia2023all}. This approach offers several distinct advantages: computational efficiency through reduced dimensionality, improved interpretability of learned representations, and seamless integration with existing language model architectures that are inherently designed for discrete token processing~\cite{du2024cosyvoice,du2024cosyvoice}. Audio tokenization, exemplified by neural codecs, has demonstrated remarkable success in compressing audio signals into compact discrete representations while maintaining reconstruction fidelity~\cite{zhang2024speechtokenizer,defossez2022high,ji2024wavtokenizer}.

The emergence of omni-modal models has introduced new architectural requirements that challenge traditional multimodal design paradigms~\cite{xu2025qwen2,wu2025step}. These systems demand the capability to process diverse multimodal inputs—including text, images, and audio—while generating high-quality discrete audio outputs for natural speech synthesis~\cite{xu2025qwen2,wu2025step,du2024cosyvoice}. This dual requirement for multimodal understanding and audio generation necessitates tokenizers that can effectively bridge continuous multimodal inputs with discrete audio representations, creating a fundamental architectural bottleneck that existing single-modal tokenizers cannot adequately address. The challenge lies in developing tokenization methods that simultaneously preserve cross-modal semantic relationships and maintain the reconstruction quality essential for natural audio generation~\cite{zhang2024speechtokenizer,zhang2025distinctive}

\subsection{Information Fusion in Multimodal Input Representations and Audio Tokenizer}

The fusion of multimodal information has been explored extensively across different representation paradigms, with varying degrees of success depending on the approach and target applications. In the domain of text-vision integration, recent work has demonstrated the effectiveness of unified tokenization strategies~\cite{zheng2024unicode,ma2025unitok}. While these approaches achieve notable results, they essentially extend discrete text representations—which already integrate seamlessly with LLM architectures—to incorporate visual information. By contrast, audio and visual modalities share fundamentally similar continuous and temporal characteristics, mirroring how humans naturally process these sensory inputs simultaneously and complementarily~\cite{meredith1986visual,knopfel2019audio}. This inherent compatibility between audio and vision suggests that their fusion represents a more natural and essential direction for multimodal tokenization, addressing a genuine representational gap rather than enhancing an already well-established text-LLM integration. 

\begin{figure*}[t]
    \centering
    \begin{subcaptionbox}{Pre-quantization fusion\label{fig:pre_fusion}}{
        \includegraphics[width=0.95\columnwidth]{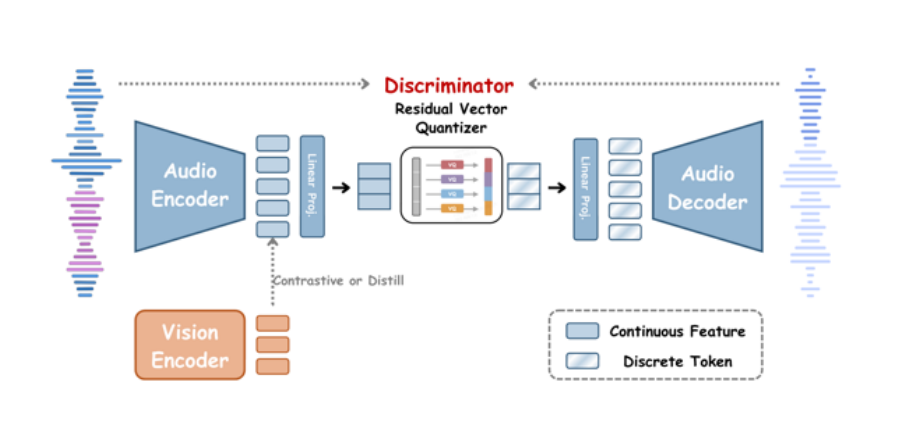}}
    \end{subcaptionbox}
    \hfill
    \begin{subcaptionbox}{Post-quantization fusion\label{fig:post_fusion}}{
        \includegraphics[width=0.95\columnwidth]{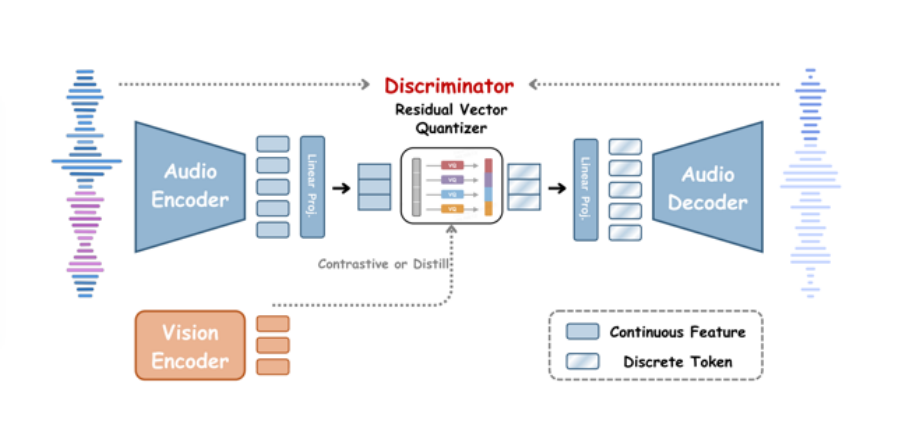}}
    \end{subcaptionbox}
    \caption{Comparison of fusion strategies in video-enhanced audio tokenization. (a) Pre-quantization fusion integrates visual features with audio representations before the quantization layer through contrastive learning or knowledge distillation. (b) Post-quantization fusion incorporates visual information within the residual vector quantizer (RVQ) during the quantization process using contrastive learning or knowledge distillation approaches.}
    \vspace{-5mm}
    \label{fig:fusion_comparison}
\end{figure*}
In continuous representation learning, contrastive methods, particularly contrastive learning~\cite{tsiamas2025sequential,li2022blip,li2023blip,xuan2023new}, have emerged as the dominant paradigm for aligning cross-modal features. While contrastive learning excels in continuous representation spaces by learning semantic alignments through positive and negative sampling, its applicability to discrete tokenizer architectures remains questionable~\cite{ma2025unitok}. The discrete nature of tokenized representations fundamentally alters the optimization landscape, potentially rendering contrastive objectives less effective or even counterproductive for tokenizer training. Given these limitations, alternative strategies such as knowledge distillation have been explored to incorporate semantic information into discrete audio tokenizers~\cite{zhang2024speechtokenizer}. These approaches typically employ semantic teachers to guide the tokenizer learning process, aiming to enhance content alignment while preserving reconstruction capabilities. However, the integration of additional semantic supervision consistently introduces a critical trade-off: although semantic information is enhanced, reconstruction quality inevitably degrades compared to purely reconstruction-focused tokenizers~\cite{zhang2024speechtokenizer,zhang2025distinctive}. This degradation directly conflicts with the high-fidelity audio generation requirements of omni-modal systems~\cite{xu2025qwen2,wu2025step}. Furthermore, extending such semantic integration approaches to truly multimodal scenarios—where audio-video information must be jointly encoded—remains largely unexplored and presents even greater challenges in balancing reconstruction fidelity with cross-modal semantic richness.

\section{Research Hypotheses and Motivating Questions}

As established in Section~\ref{sec:preliminary}, incorporating semantic or cross-modal supervision into audio tokenizers consistently degrades reconstruction quality~\cite{zhang2024speechtokenizer,gong2025xy,jiang2025unicodec}. These degradations manifest as spectral smearing, loss of high-frequency detail, temporal jitter, and weakened source separation—artifacts that compromise the fidelity required for omni-modal systems~\cite{xu2025qwen2,wu2025step,defossez2024moshi}.

We argue this degradation stems from three under-explored design choices:

\textbf{Fusion Location.} Existing methods apply cross-modal supervision at different architectural locations—within the quantization process, after quantization, or before quantization—without systematic comparison. Different locations force reconstruction and alignment gradients through different computational paths; fusion at or after the non-differentiable quantization bottleneck may cause gradient interference through biased straight-through estimators~\cite{huh2023straightening}.

\textbf{Alignment Methodology.} Current approaches adopt contrastive learning from continuous methods~\cite{radford2021learning,li2022blip}, despite discrete tokenization introducing competing quantization objectives (commit loss, codebook learning). Whether contrastive learning remains compatible with these objectives is unexplored. Knowledge distillation offers simpler feature-space supervision without competing metric learning objectives.

\textbf{Temporal Integration.} Conventional fusion enforces static frame-by-frame matching, ignoring that audio-visual events vary in salience and duration. Under token compression (e.g., 400$\rightarrow$50 tokens/sec), static fusion wastes limited capacity on uninformative regions. Dynamic temporal fusion based on content salience may better allocate representational resources.

This leads to our research question: \textbf{What architectural and optimization strategies enable cross-modal integration while preserving reconstruction fidelity?} We examine: (1) optimal fusion location, (2) compatible alignment objectives, and (3) temporal integration strategies.

We approach this through multi-objective optimization. Cross-modal tokenization balances reconstruction ($\mathcal{L}_{\text{recon}}$) and alignment ($\mathcal{L}_{\text{fusion}}$). The consistent degradation pattern suggests these objectives conflict during training, with the discrete quantization bottleneck amplifying the conflict. This motivates three hypotheses:

\textbf{Hypothesis 1: Pre-Quantization Fusion Enables Gradient Alignment.} Fusing before quantization allows both gradients to operate on continuous representations where compromise solutions exist. Fusion within/after quantization forces competing gradients through non-differentiable operations where straight-through estimators cannot reconcile conflicts.

\textbf{Hypothesis 2: Distillation Outperforms Contrastive Learning.} Contrastive learning organizes representations by cross-modal similarity while quantization organizes by reconstruction quality—these can conflict. Distillation provides direct supervision without competing constraints, enabling stable multi-objective optimization.

\textbf{Hypothesis 3: Dynamic Temporal Fusion Outperforms Static Feature Fusion.} Static fusion enforces uniform temporal correspondence, wasting capacity under compression. Dynamic fusion—varying window size based on visual salience ($\|v_t - v_{t-1}\|$)—allocates capacity to informative events, particularly critical at low token rates.

\section{Fusion Architecture Design and Methodology}

To systematically validate our hypotheses regarding optimal fusion strategies for video-enhanced audio tokenization, we design two distinct architectural approaches that differ in the location and methodology of multimodal information integration, as illustrated in Figure~\ref{fig:fusion_comparison}. Our investigation focuses on feature-dimension fusion approaches to establish rigorous experimental controls and enable fair comparison across different fusion strategies. This methodological choice allows us to isolate the effects of fusion location (pre-quantization vs. quantization-level) and fusion methodology (contrastive vs. distillation) without introducing confounding factors from different dimensional integration strategies. This also builds upon the established effectiveness of feature-dimension fusion in existing tokenizer architectures~\cite{zhang2024speechtokenizer}.

\subsection{Base Audio Tokenizer Architecture}

Our experimental framework builds upon a residual vector quantization (RVQ)~\cite{lee2022autoregressive} based audio tokenizer following the encoder-decoder paradigm, which has been widely adopted in neural audio codecs and tokenizers~\cite{zeghidour2021soundstream,defossez2022high,kumar2023high,zhang2024speechtokenizer}. The base architecture consists of three primary components: a SEANet encoder $E_{\text{audio}}$~\cite{tagliasacchi20_interspeech} that maps raw audio waveforms $x \in \mathbb{R}^{T}$ to continuous representations $z_e \in \mathbb{R}^{d \times T'}$, a residual vector quantizer $Q$ with $n_q$ quantization layers, and a SEANet decoder $D_{\text{audio}}$ for reconstruction.

The quantization process operates hierarchically across multiple layers:
\begin{equation}
\hat{z}_i = Q_i(r_{i-1}), \quad r_i = r_{i-1} - \hat{z}_i
\end{equation}
where $r_0 = z_e$ represents the initial encoder output, and each quantization layer $Q_i$ reduces the residual representation $r_{i-1}$. The final quantized representation $\hat{z} = \sum_{i=1}^{n_q} \hat{z}_i$ is passed to the decoder for reconstruction.

\subsection{Visual Feature Extraction}

Visual information is extracted using the Perception Encoder~\cite{bolya2025perception}, which provides temporally-aligned visual representations $f_{\text{vision}} \in \mathbb{R}^{d_v \times T_v}$ from input video sequences. The Perception Encoder is specifically designed for multimodal understanding tasks and produces features that capture relevant visual context for audio-visual correspondence.

\subsection{Pre-Quantization Fusion Strategy}

Our pre-quantization fusion approach integrates visual information before the quantization stage, as shown in Figure~\ref{fig:pre_fusion}. The continuous audio representation $z_e$ is enhanced with visual context through two distinct fusion methodologies:

\textbf{Knowledge Distillation Fusion}: We employ a dimension-wise distillation loss that encourages alignment between audio and visual representations:
\begin{equation}
\mathcal{L}_{\text{distill}} = -\log(\sigma(\text{cosim}(f_{\text{audio}}, f_{\text{vision}})))
\end{equation}
where $f_{\text{audio}} = \text{Transform}(z_e) \in \mathbb{R}^{T' \times d_s}$ represents the audio features projected to semantic dimension $d_s$, $f_{\text{vision}} \in \mathbb{R}^{T_v \times d_s}$ denotes the visual features, $\text{cosim}(\cdot, \cdot)$ computes cosine similarity along the feature dimension, and $\sigma$ is the sigmoid function.

\textbf{Contrastive Learning Fusion}: Following CLIP-style contrastive learning~\cite{li2022blip,ma2025unitok}, we maximize agreement between aligned audio-visual pairs:
\begin{equation}
\mathcal{L}_{\text{contrastive}} = \frac{1}{2}[\mathcal{L}_{a \rightarrow v} + \mathcal{L}_{v \rightarrow a}]
\end{equation}
where 
\begin{equation}
\mathcal{L}_{a \rightarrow v} = -\frac{1}{B}\sum_{i=1}^{B} \log \frac{\exp(s_{i,i}/\tau)}{\sum_{j=1}^{B}\exp(s_{i,j}/\tau)}
\end{equation}
and $s_{i,j} = \text{cosim}(\bar{f}_{\text{audio}}^{(i)}, \bar{f}_{\text{vision}}^{(j)})$ represents the similarity between temporally pooled features $\bar{f}_{\text{audio}} = \text{MeanPool}(f_{\text{audio}})$ and $\bar{f}_{\text{vision}} = \text{MeanPool}(f_{\text{vision}})$, with temperature parameter $\tau$.

\subsection{Quantization-Level Fusion Strategy}
Following established practices in fusion information in the audio tokenizer~\cite{zhang2024speechtokenizer}, our Quantization-Level fusion approach integrates visual information during the quantization process rather than after it, as depicted in Figure~\ref{fig:post_fusion}. Specifically, we incorporate visual features at the first quantization layer of the RVQ hierarchy, where the fusion occurs during the discrete code selection process.

Both contrastive and distillation losses are computed between the quantized features from the first layer and the corresponding visual representations. This approach enables the discrete codebook at the first quantization level to learn cross-modal correspondences while preserving the hierarchical residual structure for subsequent layers.

\subsection{Training Objectives}

The complete training objective combines reconstruction fidelity with cross-modal alignment:
\begin{equation}
\mathcal{L}_{\text{total}} = \mathcal{L}_{\text{recon}} + \lambda_{\text{mel}}\mathcal{L}_{\text{mel}} + \lambda_{\text{commit}}\mathcal{L}_{\text{commit}} + \lambda_{\text{fusion}}\mathcal{L}_{\text{fusion}}
\end{equation}

where $\mathcal{L}_{\text{recon}}$ represents L1 reconstruction loss, $\mathcal{L}_{\text{mel}}$ captures multi-scale mel-spectrogram differences, $\mathcal{L}_{\text{commit}}$ enforces commitment to discrete codes, and $\mathcal{L}_{\text{fusion}} \in \{\mathcal{L}_{\text{distill}}, \mathcal{L}_{\text{contrastive}}\}$ corresponds to the chosen fusion methodology. The weighting parameters $\lambda_{\text{mel}}$, $\lambda_{\text{commit}}$, and $\lambda_{\text{fusion}}$ balance different training objectives.

This systematic experimental design enables direct comparison between fusion locations (pre-quantization vs. quantization-level) and fusion methodologies (distillation vs. contrastive learning), providing empirical evidence for our theoretical hypotheses regarding optimal strategies to obtain multimodal discrete tokenization .

\subsection{Downstream Evaluation Framework}
Evaluating the downstream performance of discrete audio tokenizers presents fundamental methodological challenges in balancing evaluation depth with computational feasibility. Limited comparisons exist because training full-scale audio language models for evaluation requires substantial computational resources and extensive training time, making comprehensive comparison across multiple tokenization approaches prohibitively expensive. Furthermore, traditional evaluation methods that employ lightweight models on narrow tasks such as automatic speech recognition fail to capture the semantic understanding capabilities essential for multimodal applications, creating a significant evaluation gap between computational feasibility and task relevance.
\begin{figure}[t]
    \centering
    \includegraphics[width=0.5\textwidth]{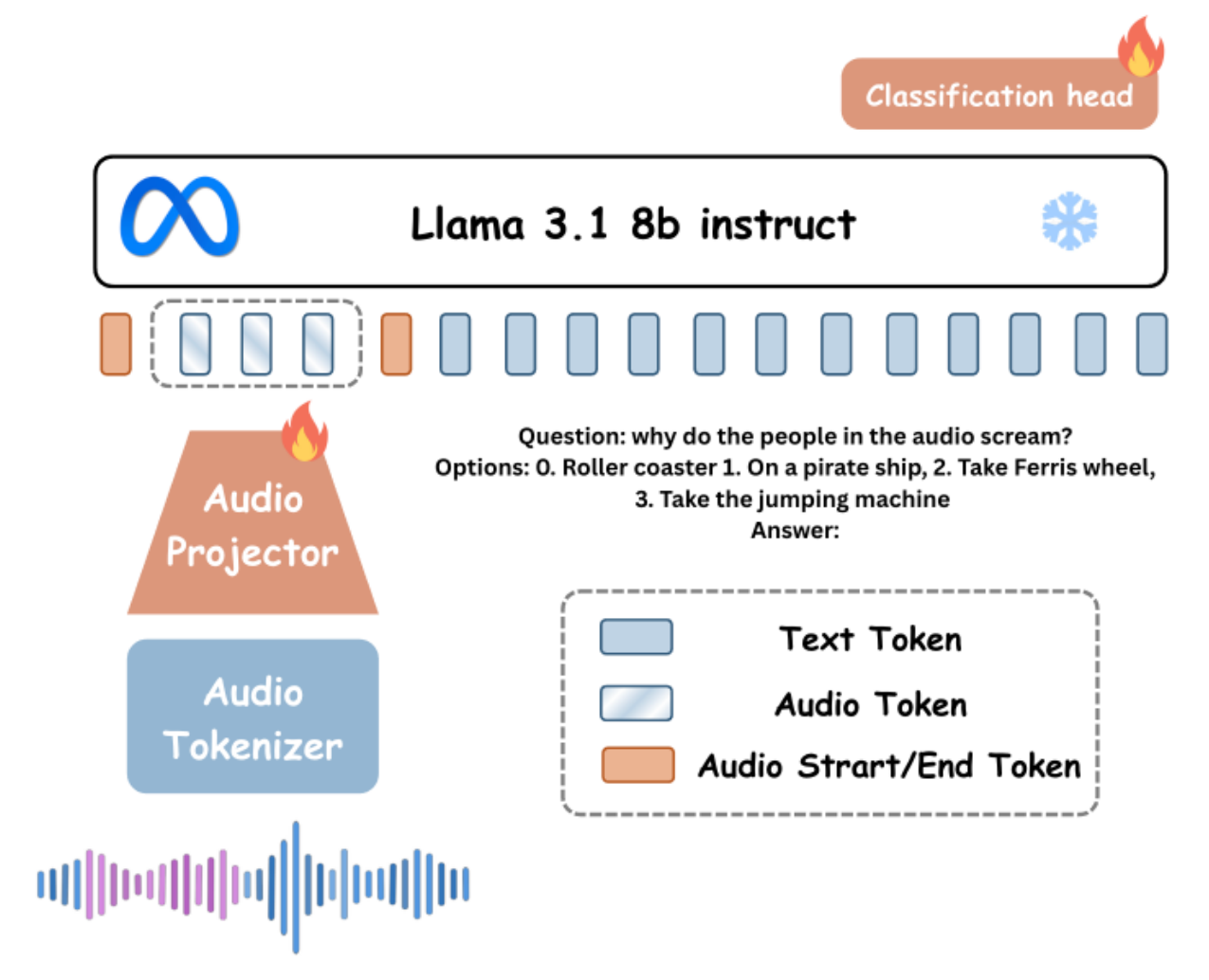}
    \caption{Downstream evaluation framework. Discrete VQ codes from the audio tokenizer are mapped to language model embeddings via a trainable Audio Projector. The Llama 3.1 8B model (frozen) processes audio tokens with textual questions for classification. Only the Audio Projector and classification head are trainable, isolating tokenization quality effects.}
    \vspace{-3mm}
    \label{fig:evaluation}
\end{figure}

To address these challenges, we adopt an evaluation methodology inspired by recent advances in discrete token projection techniques~\cite{deshmukh2023pengi}, where modifications specifically designed for systematic tokenizer comparison are implemented. Our approach leverages the Audio-Visual Question Answering (AVQA) task~\cite{sakshi2024mmau}, which provides an ideal evaluation paradigm for audio understanding capabilities due to its requirement for semantic comprehension beyond simple acoustic pattern recognition.

\textbf{Audio Token Projection Framework}: As illustrated in Figure~\ref{fig:evaluation}, our evaluation pipeline transforms discrete VQ codes into language model compatible representations through a trainable AudioProjector module. The projector employs separate embedding layers for each quantization level:
\begin{equation}
e_i = \text{Embedding}_i(c_i), \quad i \in \{1, ..., n_q\}
\end{equation}
where $c_i$ represents the VQ codes from the $i$-th quantization layer. The embeddings are concatenated and processed through a multi-layer transformer mapping network:
\begin{equation}
f_{audio} = \text{MappingNetwork}(\text{Concat}([e_1, e_2, ..., e_{n_q}]))
\end{equation}

\textbf{Classification-Based Evaluation Protocol}: Rather than employing generative evaluation approaches that introduce additional complexity, we formulate audio understanding as a multiple-choice classification task. This design choice offers simplified training procedures, more stable convergence, and direct interpretability of results. As shown in Figure~\ref{fig:evaluation}, the framework processes audio tokens alongside textual questions through a frozen language model, with only the AudioProjector and classification head remaining trainable. This protocol isolates the effects of different tokenization strategies by ensuring that performance differences can be attributed to the quality of discrete representations rather than variations in downstream model architectures or training procedures, enabling systematic comparison of pre-quantization versus quantization-level fusion approaches and contrastive versus distillation methodologies.

\section{Experimental Setup and Results}

\subsection{Experimental Configuration}

\textbf{Training Data and Model Architecture}: We train our tokenizers on AudioSet and Audioset Balance~\cite{gemmeke2017audio}, a comprehensive dataset comprising over 2 million human-labeled 10-second audio clips with 632 event classes covering diverse acoustic environments. The base tokenizer employs SEANet encoder-decoder architecture with 8-layer residual vector quantization, using 64 filters and $[8,5,4,2]$ downsampling strides for 320$\times$ compression. Models process 1024-dimensional representations through bidirectional LSTM layers, with each RVQ layer using 1024-entry codebooks.

\textbf{Training Protocol}: Models are trained for 2 epochs using AdamW optimization with learning rate $1 \times 10^{-4}$, beta parameters $\beta_1 = 0.9$, $\beta_2 = 0.99$, and effective batch size 56. Loss weighting combines reconstruction (500), commitment (10), and multi-scale mel-spectrogram losses $[45,1,1,1]$. Fusion loss weights $\lambda_{fusion} \in \{1,120\}$ follow established practices in multimodal tokenization~\cite{ma2025unitok,zhang2024speechtokenizer} to investigate reconstruction-semantic trade-offs.

\textbf{Evaluation Framework}: We assess reconstruction quality using 14,634 AudioSet test samples across four metrics: Mel Error, STFT Distance, ViSQOL~\cite{hines2015visqol}, and SI-SDR. Understanding capability is evaluated on the AVQA dataset~\cite{yang2022avqa}, containing 37,384 training and 12,528 test question-answer pairs adapted for audio-focused reasoning through keyword replacement. The downstream evaluation employs a Llama 3.1 8B model~\cite{dubey2024llama} with trainable AudioProjector and classification components, trained using AdamW (lr=$5 \times 10^{-5}$, batch size 16) for 50 epochs with bf16 mixed precision. Audio segments are processed as 30-second clips projected to 128 tokens, with classification accuracy serving as the primary understanding metric. All experiments were conducted three times with different random seeds, and we report the average results.

\subsection{Results and Analysis}
\begin{table*}[t] 
\centering
\scriptsize
\def\arraystretch{1.25}
\setlength{\tabcolsep}{4pt}
\caption{Comparison of fusion strategies for video-enhanced audio tokenization. Results show reconstruction quality metrics and downstream task performance on AVQA. Lower values indicate better performance for Mel Error and STFT Distance, while higher values are better for ViSQOL, SI-SDR, and AVQA Accuracy.}
\resizebox{0.95\textwidth}{!}{%
\begin{tabular}{@{}l@{\hspace{6pt}}c@{\hspace{6pt}}c@{\hspace{6pt}}c@{\hspace{6pt}}c@{\hspace{6pt}}c@{\hspace{6pt}}c@{\hspace{6pt}}c@{}}
\toprule[1.0pt]
\textbf{Fusion Strategy} & 
\textbf{Fusion Method} & 
\textbf{$\lambda_{fusion}$} & 
\textbf{Mel Error}$\downarrow$ & 
\textbf{STFT Distance}$\downarrow$ & 
\textbf{ViSQOL}$\uparrow$ & 
\textbf{SI-SDR}$\uparrow$ & 
\textbf{AVQA Accuracy}$\uparrow$ \\
\midrule
Audio-Only Baseline & - & 0 & 0.466 & 0.786 & 4.330 & 3.864 & 0.6474 \\
\midrule
Quantization-Level & Contrastive & 1 & 0.480 & 0.818 & 4.299 & 3.610 & 0.5399 \\
Quantization-Level & Contrastive & 120 & 0.644 & 1.173 & 3.941 & 1.215 & 0.4101 \\
Quantization-Level & Distillation & 1 &  0.481  & 0.837 & 4.248 & 3.825 & \textbf{0.6838} \\
Quantization-Level & Distillation & 120 & 0.501 & 0.869 & 4.252 & 2.775 & 0.5004 \\
\midrule
Pre-Quantization & Contrastive & 1 & 0.468 & 0.817 & 4.335 & 4.058 & 0.5507 \\
Pre-Quantization & Contrastive & 120 & 0.604 & 1.034 & 4.079 & 1.373 & 0.5685 \\
Pre-Quantization & Distillation & 1 & 0.479 & 0.825 & 4.311  & 3.258  & 0.6797 \\
Pre-Quantization & Distillation & 120 & 0.475 & 0.821 & 4.280 & 3.820 & \textbf{0.6952} \\
\bottomrule[1.0pt]
\end{tabular}%
}
\label{tab:fusion_comparison}
\end{table*}
Table~\ref{tab:fusion_comparison} presents comprehensive results comparing different fusion strategies across reconstruction quality and downstream understanding metrics. The results reveal several critical insights that validate our theoretical hypotheses and guide optimal fusion strategy selection.


\textbf{Pre-Quantization Fusion Enables Stable High-Weight Training}: The critical distinction between fusion locations lies in their behavior under varying fusion weights, not absolute performance. At $\lambda=1$, quantization-level fusion achieves competitive results (0.6838 vs 0.6797). However, increasing fusion weight reveals fundamental differences: quantization-level fusion catastrophically degrades (0.6838$\rightarrow$0.5004, -26.8\%), while pre-quantization fusion improves (0.6797$\rightarrow$0.6952, +2.3\%). This validates Hypothesis 1---pre-quantization fusion allows gradients from reconstruction and alignment to find compromise solutions in continuous space, whereas quantization-level fusion forces competing gradients through a discrete bottleneck where they interfere. Pre-quantization fusion provides a stable optimization landscape that enables strengthening multimodal supervision without triggering collapse.

\textbf{Distillation Proves Superior to Contrastive Learning in Discrete Settings}: Distillation consistently outperforms contrastive learning across both fusion locations. Most notably, contrastive learning exhibits severe degradation under high fusion weights ($\lambda=120$), with AVQA accuracy dropping to 0.4101 for quantization-level and 0.5685 for pre-quantization fusion. In contrast, distillation methods maintain stable performance, with pre-quantization distillation achieving its best results at $\lambda=120$ (0.6952 accuracy). This stark difference confirms Hypothesis 2 that continuous representation methods like contrastive learning are unsuitable for discrete tokenization settings.

\textbf{Higher Fusion Weights Benefit Distillation but Devastate Contrastive Learning}: The $\lambda$ parameter demonstrates distinct behaviors across fusion methods. For distillation approaches, higher weights ($\lambda=120$) generally improve understanding performance: pre-quantization distillation increases from 0.6797 to 0.6952 accuracy (+2.28\%), while maintaining comparable reconstruction quality (Mel Error: 0.479$\rightarrow$0.475). Conversely, contrastive learning degrades dramatically at high weights, with pre-quantization contrastive dropping from 4.058 to 1.373 SI-SDR when $\lambda$ increases from 1 to 120.

\textbf{Pre-Quantization Distillation Achieves Optimal Reconstruction-Understanding Balance}: Pre-quantization fusion with distillation at $\lambda=120$ emerges as the superior approach, achieving the highest AVQA accuracy (0.6952) while maintaining excellent reconstruction metrics (Mel Error: 0.475, ViSQOL: 4.280, SI-SDR: 3.820). This configuration outperforms the audio-only baseline by 7.38\% in understanding tasks (0.6474$\rightarrow$0.6952) while preserving competitive reconstruction quality, demonstrating successful integration of visual information without compromising tokenizer fidelity.

These comprehensive results establish three key empirical findings: (1) fusion timing before quantization is critical for preserving both reconstruction and understanding capabilities, (2) knowledge distillation provides stable and effective cross-modal alignment while contrastive learning fails catastrophically in discrete settings, and (3) careful fusion weight tuning enables substantial understanding improvements without reconstruction quality sacrifice. These discoveries validate our theoretical hypotheses and provide the empirical foundation for developing advanced temporal fusion strategies that can further exploit audio-visual correspondences in discrete tokenization frameworks.

\subsection{Gradient Analysis: Why Pre-Quantization Fusion Succeeds}

\begin{figure*}[!t]
    \centering
    \begin{subfigure}[b]{0.24\textwidth}
        \includegraphics[width=\textwidth]{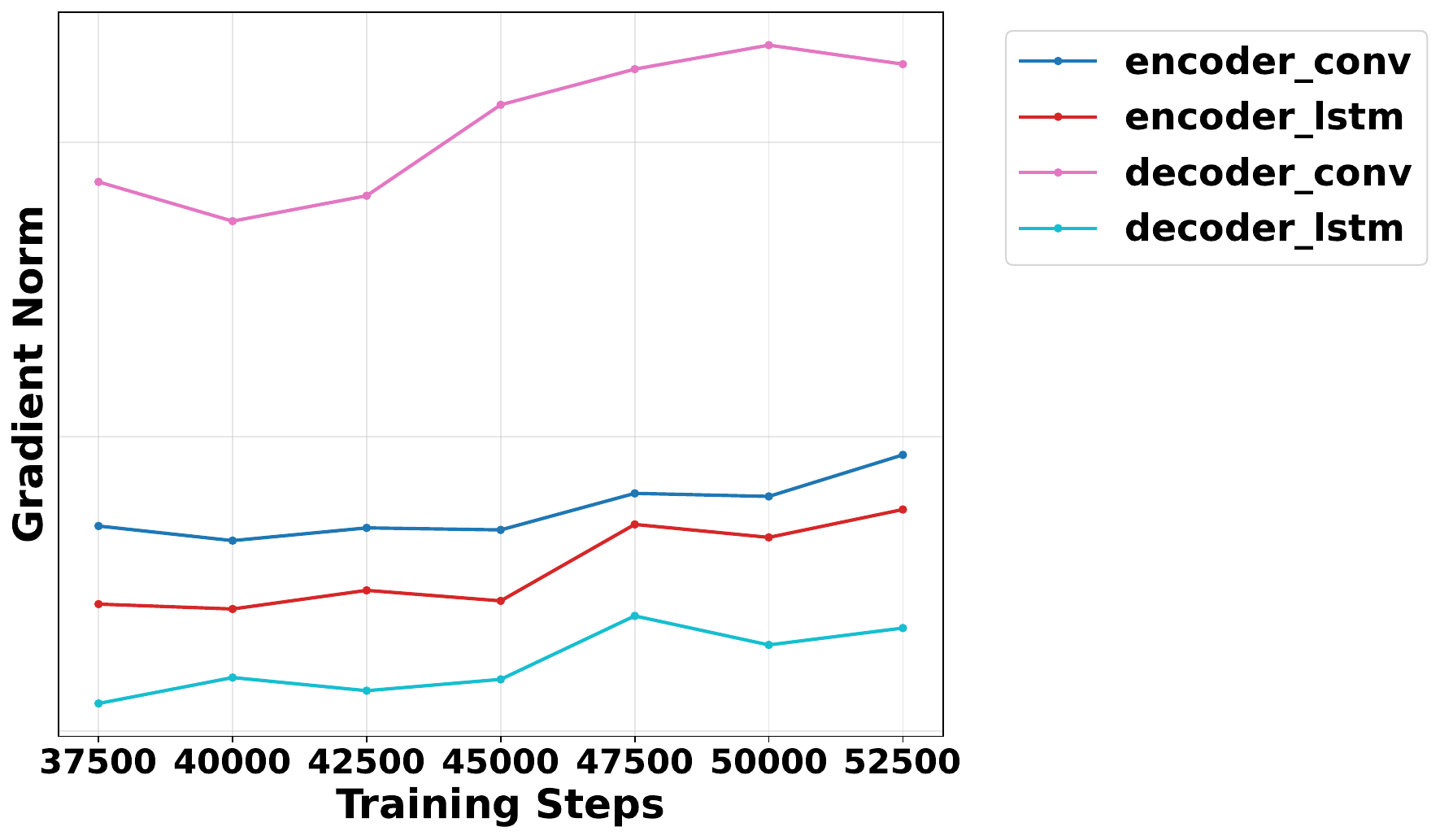}
        \caption{Gradient Evolution}
        \label{fig:grad_evo_quant}
    \end{subfigure}
    \hfill
    \begin{subfigure}[b]{0.24\textwidth}
        \includegraphics[width=\textwidth]{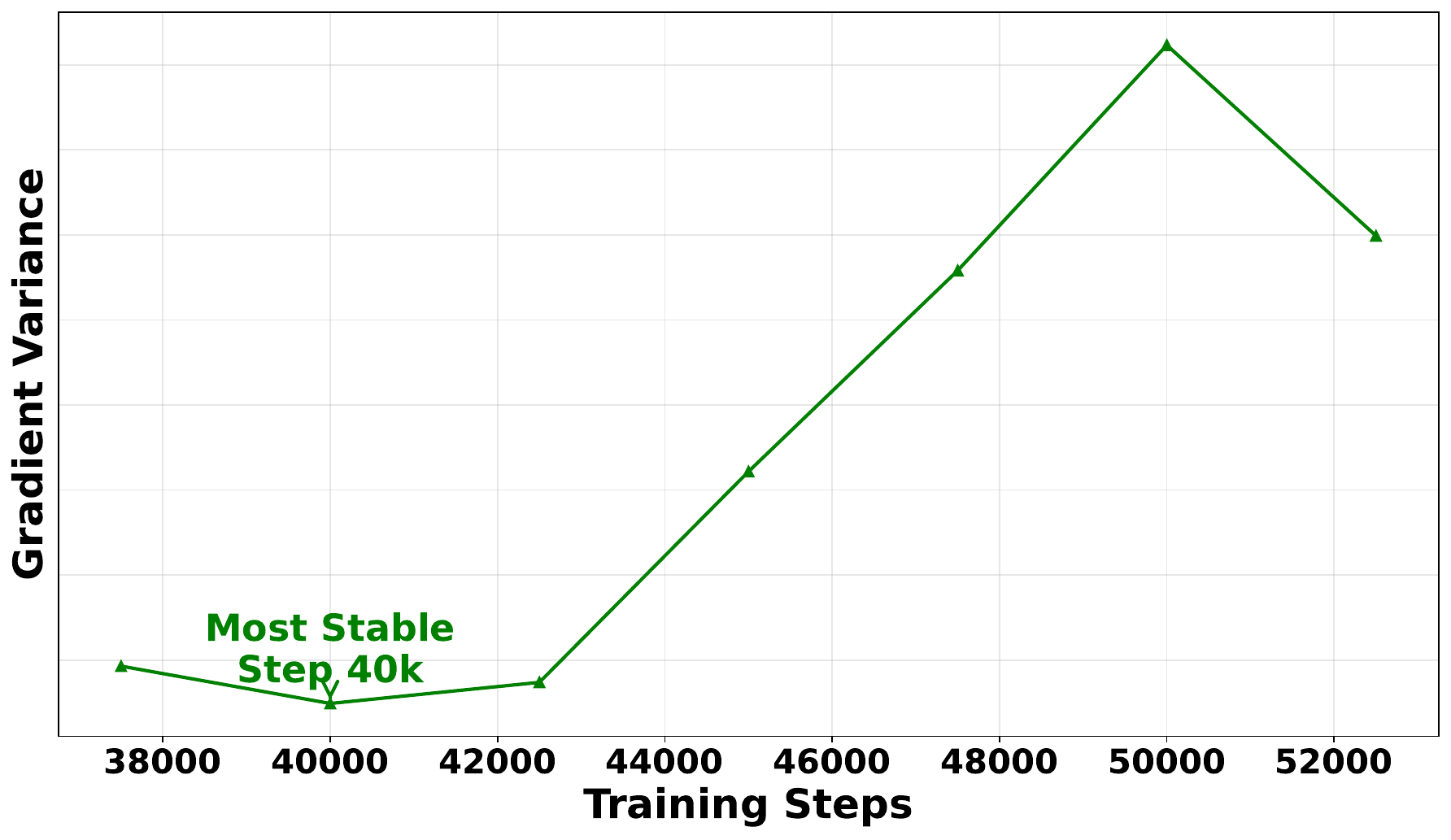}
        \caption{Gradient Stability}
        \label{fig:grad_stab_quant}
    \end{subfigure}
    \hfill
    \begin{subfigure}[b]{0.24\textwidth}
        \includegraphics[width=\textwidth]{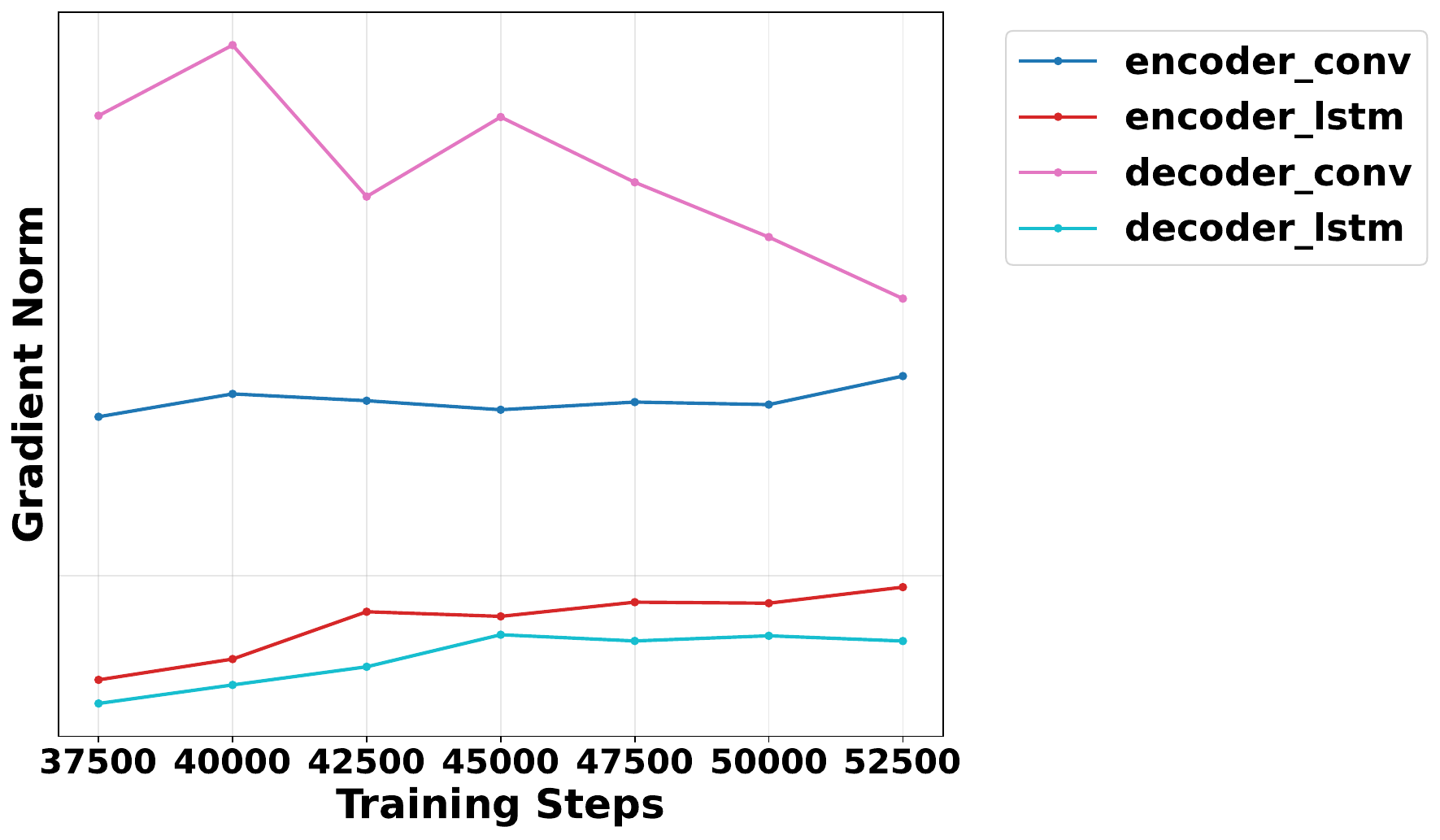}
        \caption{Gradient Evolution}
        \label{fig:grad_evo_prequant}
    \end{subfigure}
    \hfill
    \begin{subfigure}[b]{0.24\textwidth}
        \includegraphics[width=\textwidth]{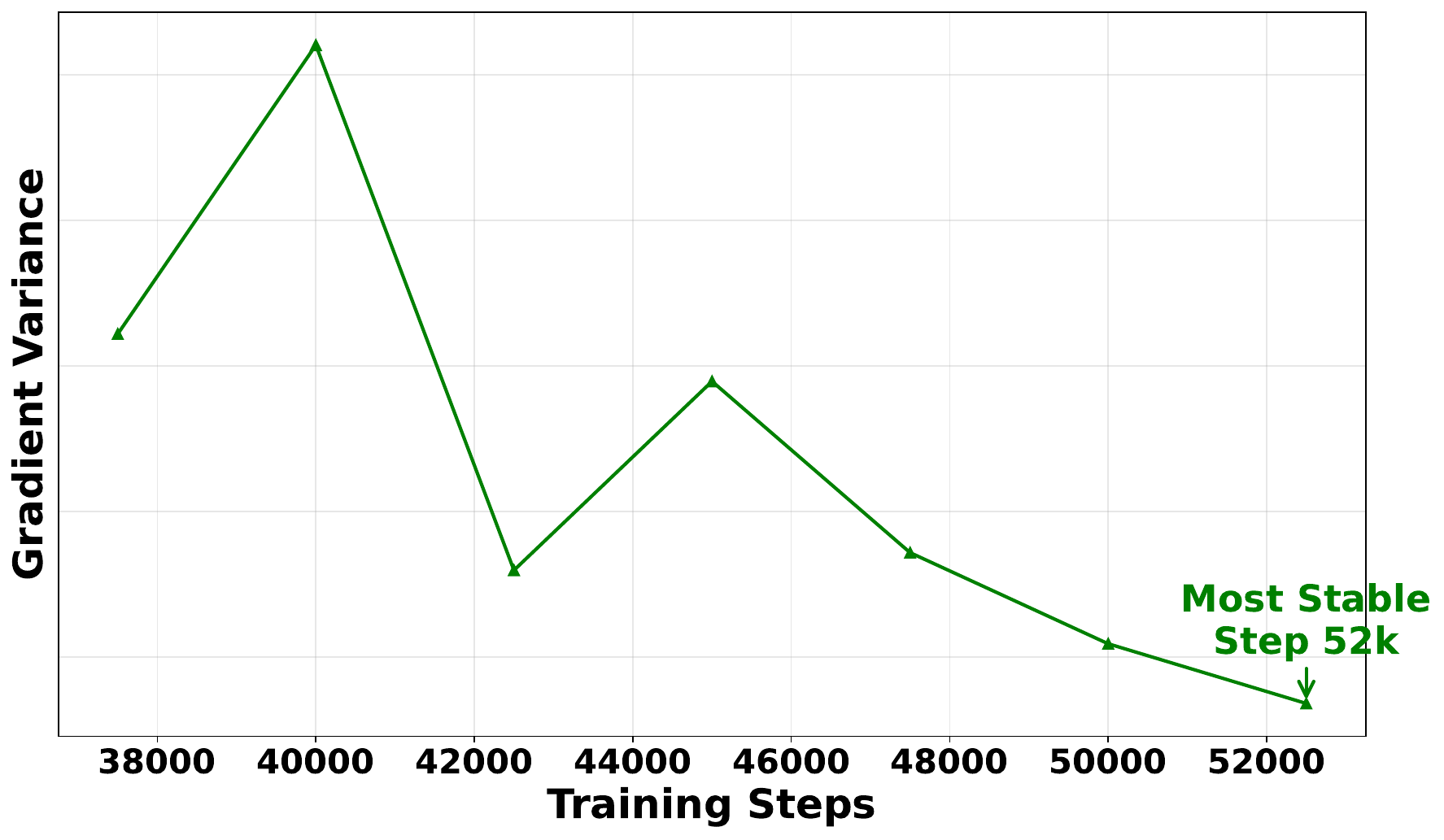}
        \caption{Gradient Stability}
        \label{fig:grad_stab_prequant}
    \end{subfigure}
    
    \vspace{0.1cm}
    
    \begin{subfigure}[b]{0.48\textwidth}
        \centering
        \textbf{Quantization-Level Fusion}
    \end{subfigure}
    \hfill
    \begin{subfigure}[b]{0.48\textwidth}
        \centering
        \textbf{Pre-Quantization Fusion}
    \end{subfigure}
    
    \caption{Gradient flow analysis comparing quantization-level fusion (left) and pre-quantization fusion (right) during training. Top row shows gradient norm evolution across model components; bottom row displays gradient variance as a stability metric. Pre-quantization fusion exhibits smoother gradient flow and higher stability, particularly in encoder and transformation layers.}
    \vspace{-5mm}
    \label{fig:gradient_analysis}
\end{figure*}
\label{sec:gradient_analysis}

Our empirical results present an optimization puzzle: why does increasing the fusion weight $\lambda$ produce opposite effects in pre-quantization versus quantization-level fusion? In the latter, raising $\lambda$ from 1 to 120 leads to a catastrophic performance drop (SI-SDR: 3.825$\rightarrow$2.775), whereas the same change in pre-quantization fusion improves understanding (AVQA accuracy: 0.6797$\rightarrow$0.6952) while preserving reconstruction quality. This divergence suggests the failure mechanism is not related to model capacity but to the optimization dynamics. We hypothesize that this problem is best understood through the lens of multi-task learning, where reconstruction and cross-modal alignment are competing objectives. Quantization-level fusion forces these objectives into conflict at the discrete bottleneck, while pre-quantization fusion enables gradient alignment in a continuous space prior to discretization.

\textbf{Methodology}. To validate this hypothesis, we analyze the gradients of the model parameters ($\theta_i$) with respect to the total loss $\mathcal{L}$ at various training checkpoints. We categorize parameters by their architectural component (e.g., encoder convolutions, decoder LSTMs). We then compute the variance of the L2-norms of these gradients, $\sigma^2_g = \text{Var}(\{\|\nabla_{\theta_i} \mathcal{L}\|\})$, which serves as a robust metric for optimization stability. In multi-objective settings, high or increasing gradient variance is indicative of gradient conflict between competing tasks, leading to optimization instability~\cite{yu2020gradient}. Conversely, decreasing variance signifies that gradients are aligning, which is characteristic of a convergent and stable optimization process~\cite{faghri2020study,keskar2017large}.

\textbf{Quantization-Level Fusion: Delayed Instability from Conflicting Gradients}.
As shown in Figure~\ref{fig:grad_evo_quant}, the gradient norms for different components in the quantization-level fusion model appear relatively stable during training. However, this masks a critical underlying issue revealed by the gradient variance analysis in Figure~\ref{fig:grad_stab_quant}. The variance remains low initially but increases sharply after 40,000 training steps. This pattern of delayed instability suggests that the conflicting gradients from the reconstruction and alignment objectives accumulate over time. The conflict is localized at the discrete bottleneck, where gradients for both objectives must pass through the non-differentiable quantization operation, typically approximated by a Straight-Through Estimator (STE)~\cite{huh2023straightening}. The STE becomes a point of severe gradient interference, as it is forced to propagate competing signals. As training progresses, the amplified alignment objective (due to a high $\lambda$) increasingly clashes with the reconstruction objective, causing the optimization to become unstable. This directly explains the performance collapse observed in Table~\ref{tab:fusion_comparison}, where high fusion weights become catastrophic for reconstruction (SI-SDR drops from 3.825 to 2.775).

\textbf{Pre-Quantization Fusion: Sustained Stability Through Gradient Alignment}.
In stark contrast, the pre-quantization fusion model exhibits fundamentally healthier optimization dynamics. Figure~\ref{fig:grad_evo_prequant} shows a convergent pattern: the initially high gradients of the decoder's convolutional layers (pink) steadily decrease, while other components maintain stable and moderate gradient norms. This visual evidence is quantitatively confirmed by Figure~\ref{fig:grad_stab_prequant}, where the gradient variance consistently decreases throughout training, indicating sustained convergence toward a stable solution. This stability is achieved because the fusion occurs in a continuous representation space \textit{before} the discrete bottleneck. Both the reconstruction gradient ($\nabla_{z_e}\mathcal{L}_{\text{recon}}$) and the alignment gradient ($\nabla_{z_e}\mathcal{L}_{\text{align}}$) are applied to the shared continuous representation $z_e$. In this space, gradient descent can effectively find a parameter update that serves as a compromise, jointly satisfying both objectives. The quantizer then simply discretizes this already-optimized representation. This robust optimization process persists even at high fusion weights, which explains why stronger multimodal supervision successfully enhances understanding (AVQA accuracy improves to 0.6952) without compromising reconstruction quality.

\textbf{Explaining the Dichotomy of Fusion Weight}.
This gradient analysis provides a clear resolution to our initial puzzle.
\begin{itemize}
    \item In \textbf{quantization-level fusion}, increasing $\lambda$ amplifies the alignment objective, which intensifies the gradient interference at the non-differentiable bottleneck and accelerates optimization instability.
    \item In \textbf{pre-quantization fusion}, increasing $\lambda$ strengthens the alignment objective in a continuous space where gradients can be effectively combined and aligned, thus improving cross-modal understanding without disrupting the overall convergence.
\end{itemize}
Our optimization-based hypothesis is therefore validated: pre-quantization fusion succeeds by \textbf{separating the multi-objective optimization from the discretization step}. In contrast, quantization-level fusion fails by forcing competing objectives through a discrete bottleneck where their gradients conflict. This mechanism offers a unified explanation for why previous semantic enhancement approaches that operate at or after the quantization layer have consistently resulted in degraded audio reconstruction quality~\cite{zhang2024speechtokenizer,gong2025xy,jiang2025unicodec}.

\section{Timing-Aware Pre-Quantization Fusion for Video-Enhanced Audio Tokenization}
\label{sec:tapf_main}
\begin{figure*}[t]
    \centering
    \includegraphics[width=1\textwidth]{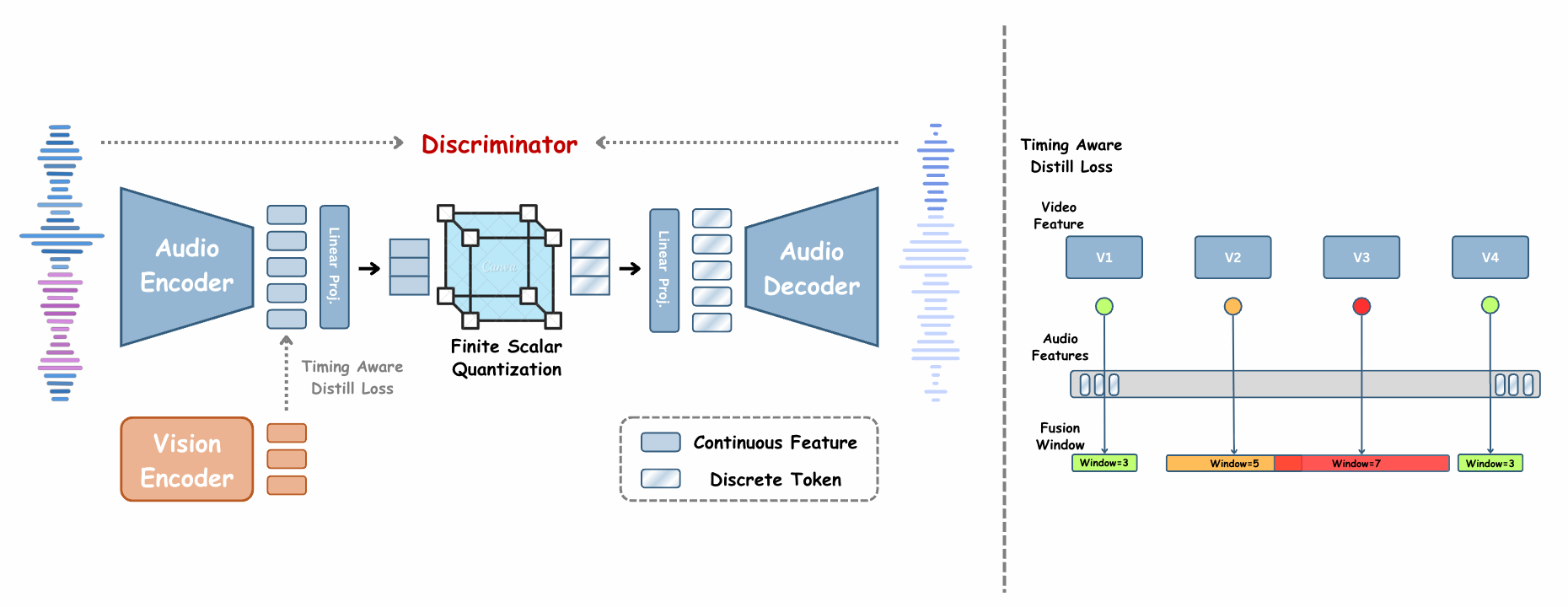}
    \caption{Overview of the proposed Timing-Aware Pre-Quantization Fusion (TAPF) architecture. Left: Visual features are fused with continuous audio features before quantization, guided by a novel distillation loss. Right: The loss mechanism aligns each video feature with audio features inside a dynamically-sized temporal window, enabling flexible cross-modal alignment.} 
    \vspace{-6mm}
    \label{fig:pipeline}
\end{figure*}

Our previous experiments established two critical findings: (1) integrating cross-modal information \textit{before} the quantization bottleneck is essential for preserving reconstruction fidelity, and (2) knowledge distillation is a more stable and effective alignment mechanism than contrastive learning in discrete tokenizer architectures. However, these feature-dimension fusion strategies still possess a fundamental limitation: they assume a static, one-to-one temporal correspondence between audio and visual features. This rigid assumption fails to capture the complex, dynamic nature of real-world audio-visual events, where the duration and salience of an audio event may not align with a fixed-length visual segment. To overcome this, we propose that a truly effective fusion strategy must operate along the temporal axis, dynamically adapting its focus based on the informational content of the modalities.

Drawing inspiration from \textbf{Distinctive Feature Theory}~\cite{lee1988automatic,liu1996landmark}, which posits that the most critical information in a speech signal is concentrated around moments of significant acoustic change (i.e., landmarks), we extend this principle to the multimodal domain. We hypothesize that moments of significant \textit{visual} change are temporally correlated with the most salient audio events. For instance, the sound of a shattering glass corresponds directly to the visual event of the glass breaking. Based on this, we introduce \textbf{Timing-Aware Pre-Quantization Fusion (TAPF)}, a novel method that replaces static feature alignment with a dynamic, content-aware temporal pooling mechanism.

\subsection{Dynamic Temporal Alignment Mechanism}
The architecture of TAPF is illustrated in Figure~\ref{fig:pipeline}. The left panel shows the overall framework implementing pre-quantization fusion, as established in Section V. The core innovation is the \textbf{Timing-Aware Distill Loss} mechanism detailed in the right panel. We primarily employ Finite Scalar Quantization rather than RVQ, as FSQ's lower token rate (50 vs 400 tokens/sec) enables tractable downstream LLM training while maximizing the benefits of dynamic fusion under token scarcity.

Let the sequence of continuous audio features from the encoder be $Z_e = (z_1, z_2, \dots, z_{T'}) \in \mathbb{R}^{T' \times d}$ and the sequence of video features be $V = (v_1, v_2, \dots, v_{T_v}) \in \mathbb{R}^{T_v \times d_v}$. The TAPF mechanism operates as follows:

\textbf{1) Visual Salience as a Proxy for Change:} For each video feature frame $v_t$, we compute a visual complexity score $c_t = \|v_t - v_{t-1}\|_2$, serving as a proxy for visual change or salience. A higher score indicates more significant events. For $t=1$, we use $c_1 = \|v_1\|_2$.

\textbf{2) Dynamic Temporal Windowing:} The complexity score dynamically determines the temporal window size $W_t$ over the audio feature sequence:
\begin{equation}
    W_t = \text{round}(W_{min} + (W_{max} - W_{min}) \cdot \sigma(c_t))
\end{equation}
where $\sigma(\cdot)$ is the sigmoid function. We set time windows from $W_{min}=1$ to $W_{max}=7$, allowing adaptive focus from localized to broader temporal regions. Low complexity yields narrow windows for precise alignment; high complexity produces wider windows capturing broader audio context corresponding to salient visual events.

\textbf{3) Attention-based Temporal Pooling:} Within the window centered at the corresponding audio time step, we identify audio feature indices $\mathcal{J}_t$ and compute attention weights based on cosine similarity:
\begin{equation}
    \alpha_{t,j} = \frac{\exp(\text{cosim}(v_t, z_j))}{\sum_{k \in \mathcal{J}_t} \exp(\text{cosim}(v_t, z_k))}, \quad \hat{z}_t = \sum_{j \in \mathcal{J}_t} \alpha_{t,j} z_j
\end{equation}
This aggregates the most relevant audio features within each variable-length segment.

The Timing-Aware Distill Loss combines L1 and cosine similarity losses:
\begin{equation}
    \mathcal{L}_{TAPF} = \frac{1}{T_v}\sum_{t=1}^{T_v} \left( \|\hat{z}_t - v_t\|_1 + \lambda_{sim}(1 - \text{cosim}(\hat{z}_t, v_t)) \right)
\end{equation}
where $\lambda_{sim}=1.0$ balances feature matching and angular similarity.

\subsection{Main Results and Analysis}

The performance of our proposed TAPF model is evaluated against relevant baselines in Table~\ref{tab:main_results_final}.

\textbf{Dynamic Fusion Provides Negligible Gains at High Token Rates but Substantial Improvements Under Compression.} At 400 tokens/sec (RVQ8), TAPF improves AVQA accuracy by 3.68\% over A-V Static Fusion (0.7208 vs 0.6952). At 50 tokens/sec (FSQ), the improvement increases to 19.0\% (0.6941 vs 0.5832). At high token rates, dense representations minimize quantization losses regardless of fusion strategy. Under aggressive compression (50 tokens/sec), static fusion's uniform temporal sampling wastes token capacity on redundant regions, while TAPF concentrates tokens on high-salience moments.

\textbf{Dynamic Alignment Simultaneously Improves Both Reconstruction and Understanding.} At 50 tokens/sec, TAPF achieves 3.9\% higher reconstruction quality than A-V Static Fusion (ViSQOL: 4.097 vs 3.942) while simultaneously gaining 19.0\% in understanding performance. Section V-C's gradient analysis explains this: pre-quantization fusion allows the codebook to optimize for acoustic fidelity without conflicting cross-modal alignment objectives.

\textbf{Multimodal Fusion Achieves 8× Token Efficiency Over Audio-Only Models.} TAPF outperforms WavTokenizer by 7.0\% (0.7208 vs 0.6734) despite similar architectural complexity. TAPF at 50 tokens/sec (0.6941) matches audio-only models operating at 400 tokens/sec, demonstrating 8× compression efficiency. Speech Tokenizer's lower performance (0.5839) confirms that domain-specific training cannot replicate multimodal integration benefits.

\textbf{TAPF Achieves Asymmetric Trade-off Favoring Understanding Over Reconstruction Loss.} Previous methods degraded reconstruction when enhancing semantics~\cite{zhang2024speechtokenizer}. Compared to the audio-only baseline (ViSQOL: 4.330, AVQA: 0.6474), TAPF loses 0.5\% reconstruction quality while gaining 11.3\% understanding performance—a 22:1 benefit-to-cost ratio. Pre-quantization fusion with content-aware alignment separates multimodal objectives from discrete quantization constraints.
\begin{table}[t]
\centering
\caption{Performance comparison of the proposed TAPF model against audio-visual and audio-only baselines on reconstruction quality (ViSQOL) and downstream understanding (AVQA Accuracy).}
\label{tab:main_results_final}
\scriptsize
\def\arraystretch{1.2}
\setlength{\tabcolsep}{1pt}
\resizebox{0.99\columnwidth}{!}{
\begin{tabular}{lcccccc}
\toprule
\textbf{Model} & \textbf{Dataset} & \textbf{Quantizer} & \textbf{Frame Rate} & \textbf{Token Rate} & \textbf{ViSQOL $\uparrow$} & \textbf{AVQA Accuracy $\uparrow$} \\
\midrule
\multicolumn{7}{l}{\textit{Audio-Visual Baseline}} \\
A-V Static Fusion & AudioSet & RVQ8 & 50 & 400 & 4.280 & 0.6952 \\
A-V Static Fusion & AudioSet & FSQ & 50 & 50 & 3.942 & 0.5832 \\
\midrule
\multicolumn{7}{l}{\textit{Audio-Only Tokenizer Baselines}} \\
WavTokenizer & Several & VQ & 75 & 75 & 4.332 & 0.6734 \\
DAC Tokenizer & Several & RVQ9 & 31.25 & 281 & 4.304 & 0.6561 \\
Speech Tokenizer & LibriSpeech & RVQ8 & 50 & 400 & 4.129 & 0.5839 \\
\midrule
\multicolumn{7}{l}{\textit{Proposed Audio-Visual Model}} \\
\textbf{TAPF} & AudioSet & RVQ8 & 50 & 400 & 4.308 & \textbf{0.7208} \\
TAPF & AudioSet & FSQ & 50 & 50 & 4.097 & 0.6941 \\
\bottomrule
\vspace{-6mm}
\end{tabular}
}
\end{table}

\subsection{Ablation Study for TAPF}
We systematically evaluate each TAPF component through controlled ablations (Table~\ref{tab:fsq_ablation_compact}), revealing the hierarchical contribution of each design choice to the final performance.

\textbf{Dynamic Windowing is the Critical Enabler, Not Just an Enhancement.} Removing dynamic windowing causes catastrophic understanding collapse (0.6941$\rightarrow$0.5160, -25.6\%) with minimal reconstruction impact (-2.4\%). This stark asymmetry reveals a fundamental principle: \emph{the token allocation problem dominates the representation quality problem under compression}. Fixed-length windows force uniform distribution of limited token capacity across all temporal regions, effectively averaging out salient events. The fact that reconstruction quality barely suffers (-2.4\%) while understanding collapses (-25.6\%) demonstrates that the audio signal itself remains adequately preserved---the failure is in \emph{where} the model focuses its limited representational budget.

\textbf{Window Size Operates in a Narrow Optimal Range.} Expanding the maximum window size from 5 to 7 yields substantial gains (0.4900$\rightarrow$0.6941, +41.7\%), but further expansion to 9 provides diminishing returns (0.6941$\rightarrow$0.6903, -0.5\%). This suggests an intrinsic temporal scale of audio-visual correspondence: windows must be large enough to capture the temporal extent of typical events (requiring $W_{max} \geq 7$), but excessively large windows reintroduce the averaging problem that dynamic windowing was designed to solve. The optimal range ($W_{max} = 7$, corresponding to $\sim$140ms at 50 fps) aligns with psychoacoustic findings on the temporal integration window for audio-visual binding.

\textbf{Visual Complexity Metric Matters Less Than Expected.} The choice between L1 and L2 norm for measuring visual change shows minimal impact (0.6891 vs 0.6941). This robustness suggests that TAPF's effectiveness stems from the \emph{relative} salience signal rather than absolute magnitude precision. Any reasonable change detector suffices as long as it provides monotonic ordering of temporal importance---the attention mechanism and window sizing collectively handle the rest.

\begin{table}[t]
\centering
\caption{Ablation Study of TAPF Components (FSQ Quantizer)}
\label{tab:fsq_ablation_compact}
\renewcommand{\arraystretch}{1.2}
\setlength{\tabcolsep}{4pt}
\begin{tabular}{l c c}
\toprule
\textbf{Model Configuration} & \textbf{ViSQOL $\uparrow$} & \textbf{AVQA Acc. $\uparrow$} \\
\midrule
\textbf{TAPF (Full Model)}    & \textbf{4.097} & \textbf{0.6941} \\
\midrule
\multicolumn{3}{l}{\textit{Ablation of Core Components}} \\
- w/o Dynamic Window          & 3.997 & 0.5160 \\  
\midrule
\multicolumn{3}{l}{\textit{Window Size Variants}} \\
$W_{\text{min}}=1, W_{\text{max}}=5$ & \textcolor{black}{3.98} & \textcolor{black}{0.4900} \\
$W_{\text{min}}=1, W_{\text{max}}=7$ (Ours) & \textbf{4.097} & \textbf{0.6941} \\  
$W_{\text{min}}=1, W_{\text{max}}=9$ & \textcolor{black}{3.93} & \textcolor{black}{0.6903} \\
\midrule
\multicolumn{3}{l}{\textit{Visual Complexity Metrics} } \\
L1 Norm         & \textcolor{black}{4.043} & \textcolor{black}{0.6891} \\
L2 Norm (Ours)                & \textbf{4.097} & \textbf{0.6941} \\  %
\midrule
\multicolumn{3}{l}{\textit{Pooling Strategies}} \\
Mean Pooling                  & 4.011 & 0.5889 \\
Attention Pooling (Ours)      & \textbf{4.097} & \textbf{0.6941} \\  
\bottomrule
\vspace{-7mm}
\end{tabular}
\end{table}

\textbf{Attention Provides Necessary Local Refinement.} Replacing attention pooling with mean pooling degrades understanding by 15.1\% (0.6941$\rightarrow$0.5889) while minimally affecting reconstruction. This clarifies the architectural hierarchy: \emph{dynamic windowing solves the global resource allocation problem (which regions matter), while attention solves the local aggregation problem (how to combine features within each region)}. Mean pooling treats all features within a window equally, discarding fine-grained correspondence cues that attention preserves. Notably, attention-only ablation (0.5889) outperforms static fusion (0.5832), confirming that even without dynamic windows, learned attention weights better exploit audio-visual correspondence than fixed uniform sampling.

\textbf{The Components Form a Multiplicative, Not Additive, System.} The full TAPF model (0.6941) substantially outperforms any single-component ablation, yet the performance cannot be explained by simple addition of individual contributions. Dynamic windowing alone achieves only 0.5160, and attention alone reaches 0.5889, but their combination yields 0.6941---a synergistic effect. This multiplicative interaction validates our design philosophy: \emph{content-aware temporal allocation and fine-grained local alignment must operate in concert to maximize information capture under severe token constraints}. Neither component alone addresses both the global and local aspects of the multimodal fusion problem.

\section{Conclusion}
This work resolves the apparent incompatibility between semantic enhancement and reconstruction fidelity in multimodal audio tokenization. Through systematic investigation, we demonstrate this tension stems from architectural choices forcing competing objectives through discrete bottlenecks where gradients conflict. Our investigation establishes three principles. First, \emph{optimization topology matters more than fusion complexity}---pre-quantization fusion enables finding compromise solutions in continuous space before discretization. Second, \emph{stability dominates expressiveness}---contrastive learning fails catastrophically in tokenizers, while distillation maintains convergence by avoiding objective interference. Third, \emph{token scarcity transforms fusion into resource allocation}---under compression, \emph{where} to allocate capacity outweighs \emph{how} features combine. The proposed approach, TAPF, operationalizes these principles through content-aware temporal allocation, demonstrating that compression fundamentally changes multimodal fusion and suggesting future work should prioritize principled resource allocation over sophisticated fusion mechanisms.

\bibliographystyle{IEEEtran}

\begin{footnotesize}
\bibliography{reference}
\end{footnotesize}

\end{document}